\documentclass[showpacs,preprintnumbers,amsmath,reprint,aps,prl,superscriptaddress,twocolumn]{revtex4-2}

\usepackage{bm}
\usepackage{graphicx}
\usepackage{color}
\usepackage{amsmath}
\usepackage{hyperref}
\usepackage{epstopdf}
\usepackage{upgreek}
\usepackage{mathptmx, textcomp}
\usepackage[latin1]{inputenc}
\usepackage[T1]{fontenc}
\usepackage{array, multirow}
\usepackage[table]{xcolor}
\usepackage{booktabs}
\usepackage{longtable}

\usepackage[normalem]{ulem}

\newcommand{\commentOut}[1]{}

\begin{document}

\title{Cold trapped molecular ions and hybrid platforms for ions and neutral particles}

\author{Markus Dei{\ss}}
\affiliation{Institut f\"{u}r Quantenmaterie and Center for Integrated Quantum Science and Technology IQ$^{ST}$, Universit\"{a}t Ulm, D-89069 Ulm, Germany}

\author{Stefan Willitsch}
\affiliation{Department of Chemistry, University of Basel, 4056 Basel, Switzerland}

\author{Johannes Hecker Denschlag}
\affiliation{Institut f\"{u}r Quantenmaterie and Center for Integrated Quantum Science and Technology IQ$^{ST}$, Universit\"{a}t Ulm, D-89069 Ulm, Germany}

\date{\today}

\begin{abstract}
We review recent progress in the field of cold trapped molecular ions. A new generation of collision and cold chemistry experiments between atoms and ions has emerged,
where cold atoms and ions are brought into contact in a controlled way in novel 
hybrid atom-ion platforms. Furthermore, new possibilities for the preparation and detection of molecular quantum states with high sensitivity and precision have been demonstrated based on quantum-logic schemes. These advances represent important stepping stones for new directions in fundamental research and technological applications across various domains including precision measurements, quantum technologies and chemical dynamics.
\end{abstract}

\maketitle

\section{Introduction}

Over the past few decades, advances in the realm of atomic physics have paved the way to new research directions and technologies such as ultracold quantum gases, extremely precise optical atomic clocks and state-of-the-art platforms for quantum information. These spectacular successes have kindled a renewed interest in quantum experiments on molecular systems. Molecules have more degrees of freedom than atoms, rendering experiments considerably more challenging, but this increased complexity also constitutes a powerful resource for new applications in science and technology.

\begin{figure}[t]
	\includegraphics[width=\columnwidth]{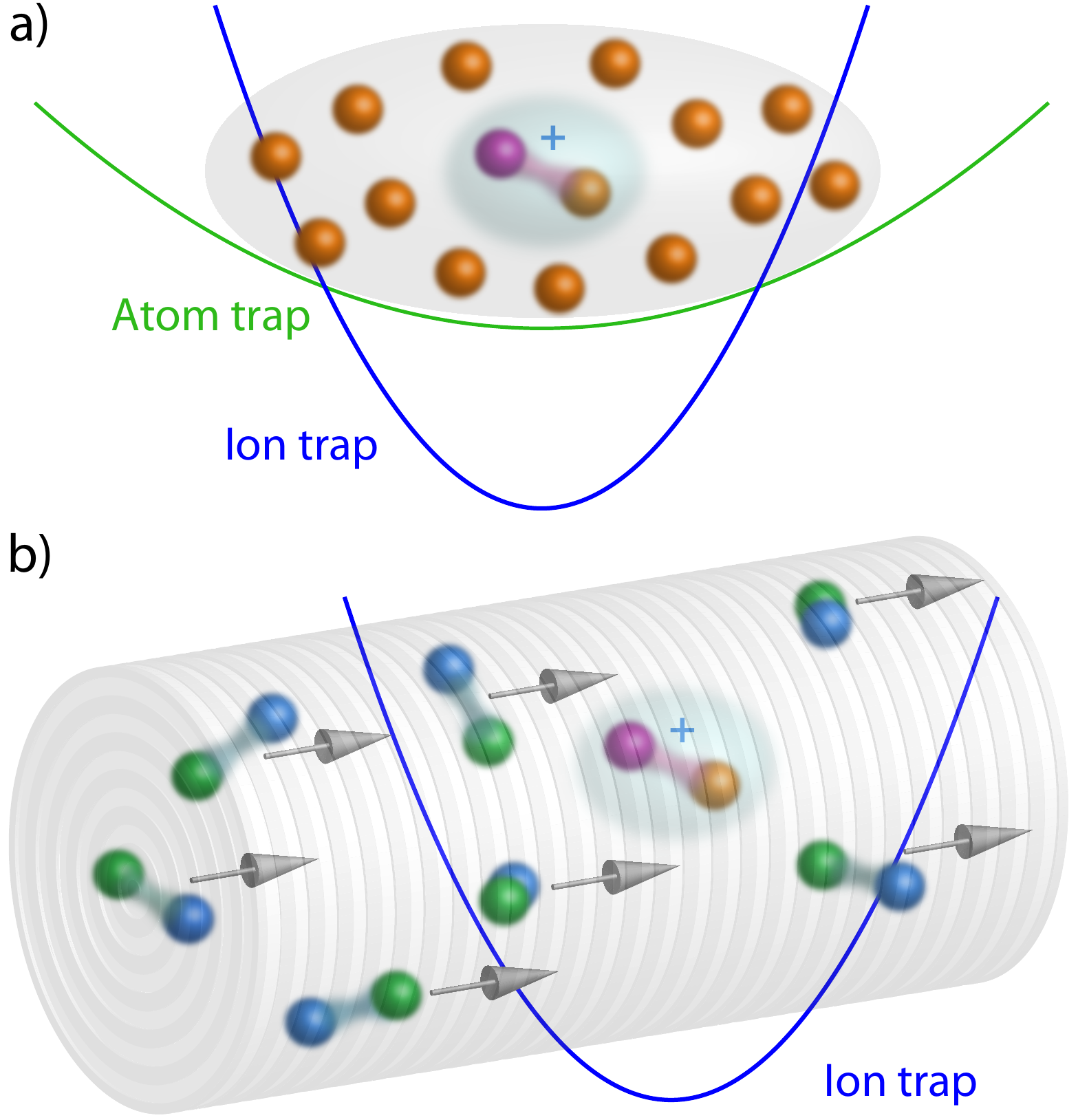}
	\caption{Hybrid platforms for ions and neutral particles. (a) Combined atom-ion trap setup. (b) Combination of an ion trap with a molecular beam.
	}
	\label{Fig1}
\end{figure}

Molecular ions represent an important species, since, due to their charged nature, they offer specific opportunities for trapping and probing. In recent years, powerful methods have been developed for the preparation of molecular ions in well-defined quantum states and for studies of their interactions with their environment on the quantum level.
In particular, hybrid platforms have been implemented which integrate cold trapped ions with cold neutral particles \cite{Haerter2014,Willitsch2012,Lous2022,Tomza2019}. Typical experimental setups combine an ion trap with either an atom trap or a directed atomic or molecular beam (see Figs. \ref{Fig1} (a) and \ref{Fig1} (b), respectively).
These developments open up exciting new possibilities for fundamental research as well as for applications in fields such as precision metrology, quantum sensing, cold chemistry, and quantum information.

Up until about 15 years ago, the fields of ion trapping \cite{Leibfried2003,Haffner2008} and ultracold neutral atomic gases \cite{Bloch2008} were largely separated, each developing their own distinct experimental tools. In ion traps, typically samples of one to a few ions are prepared which can be individually manipulated and detected. The ion trap is, in general, sufficiently deep so that an ion is not lost, even after a strongly exothermal chemical reaction. Cold neutral atoms can typically be produced in large numbers, high densities and at extremely low temperatures (down to nK), and they can form beams with well defined velocities and directions. Similarly, the technology for the generation and manipulation of cold neutral molecules has advanced rapidly over the past years \cite{Quemener2012,Softley2023}. The combination of the technologies for neutrals and ions generates platforms in which cold collisions and cold chemistry between individual atoms and ions can be studied in a precisely controlled environment. In particular, ion and neutral atom traps do not disturb each other significantly and can be separated or merged on demand. Furthermore, at low enough temperatures one can potentially control the interactions between the particles with the help of magnetically tunable Feshbach resonances \cite{Chin2010}. Achieving such temperatures has, however, been challenging due to heating effects in ion traps. Nonetheless, reaching this regime has recently been demonstrated for specific atom-ion systems and Feshbach resonances have been observed. 

Besides the advancement of hybrid platforms, additional important technological progress has taken place in recent years. For example, novel non-destructive detection schemes based on quantum-logic protocols have been developed for probing the quantum states of molecular ions. These schemes can also be used for the state-selective preparation of the molecules and for highly sensitive spectroscopy.

Building on this technological basis, a variety of lines of research with molecular ions 
have been developed. These include new approaches for the production of cold molecular ions as well as studies of reactive and inelastic processes with molecular species. For such studies, state-specificity and control aspects are of particular interest. Finally, impressive progress has been achieved in the precision spectroscopy of molecular ions, see e.g. \cite{Roth2008,Safronova2018}. Here, fundamental physical parameters and constants are measured, and theoretical models, such as the standard model, can be tested with ever higher accuracy.
  
In the following, we give an overview of recent research highlights and scientific developments involving cold molecular ions. We will first discuss methods for the generation of molecular ions in specific motional and internal quantum states and for their state detection. Then, we will give a short introduction to the radiofrequency ion trap (Paul trap) as the central enabling technology of the field and to the ion micromotion occurring in these traps which is the main obstacle for reaching the $s$-wave collisional regime. After recounting important properties of atom-ion interaction and collisions in a section, we dedicate a large part to an overview of chemical reactions and inelastic processes. Finally we will discuss recent progress in precision spectroscopy before we will conclude the review by highlighting some promising directions for future work. 

In this article, we focus on experiments in the "cold" regime at temperatures below 1~K. For a discussion on research with molecular ions at temperatures down to 10~K, which are of relevance for, e.g., astrophysics, the reader is referred to, e.g., Refs. \cite{McGuire2020,Wester2009,Gerlich1992b} and references therein.

\section{Preparation, cooling and detection of molecular ions}

\begin{figure}[t]
	\includegraphics[width=\columnwidth]{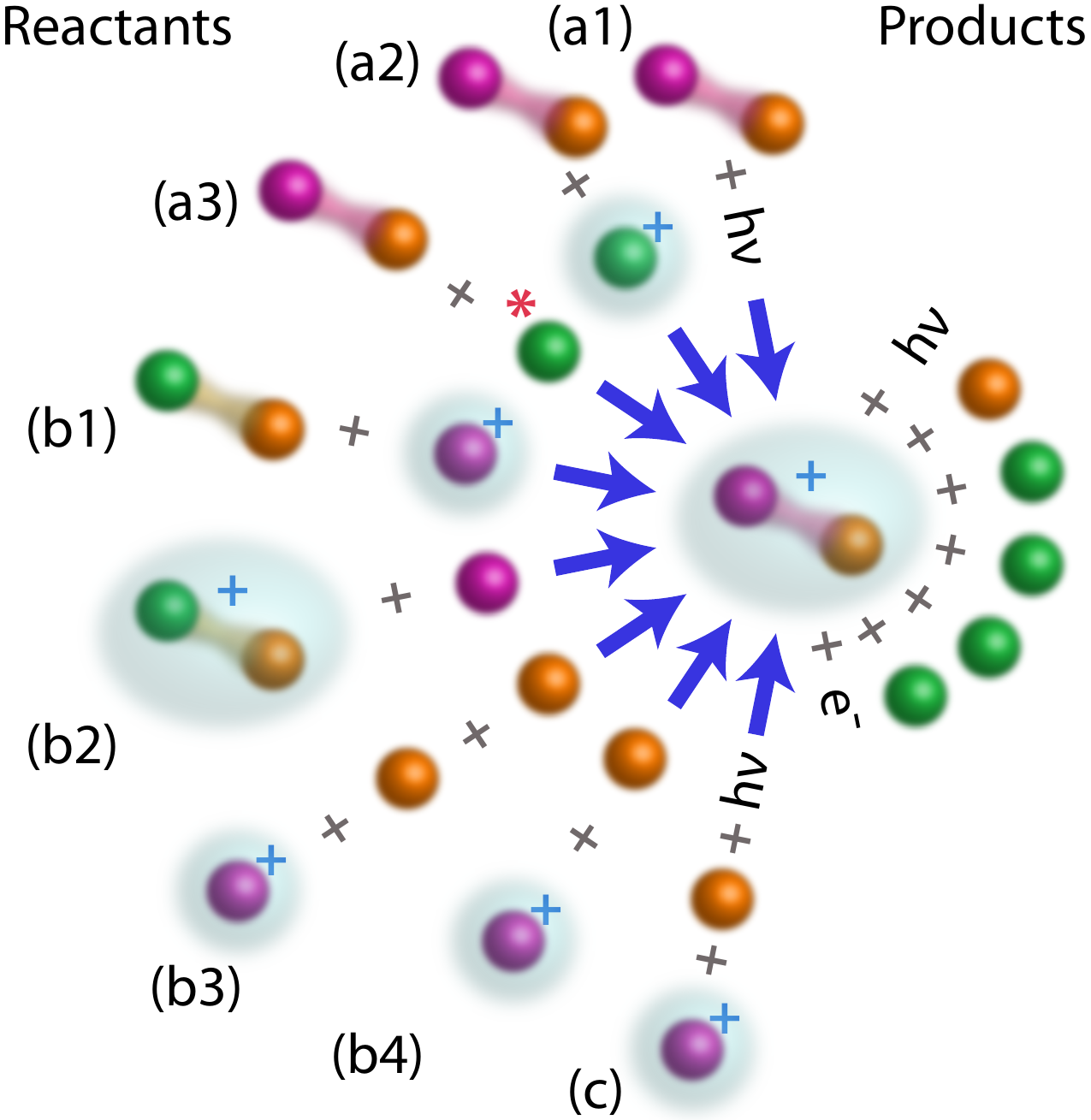}
	\caption{Several formation processes of a molecular cation. The corresponding reaction equations  are arranged in a circular fashion, sharing the formed molecular ion in the centre. Group (a) includes schemes for the ionisation of neutral precursor molecular species. (a1): Photoionisation; (a2): Charge transfer; (a3): Penning ionisation, where the neutral molecule encounters an internally excited collision partner as indicated by the red asterisk resulting in molecular ion formation while the collision partner is deexcited. In group (b), examples of reactive processes are given. In this category, we consider collisions of reactants that lead to the formation, the disintegration, or the change of one or several chemical bonds between the atomic constituents involving the production of a molecular ion. (b1,b2): Substitution reactions; (b3): Atom-atom-ion three-body recombination; (b4): Radiative association. Finally, in photoassociation (c) the absorption of a photon takes place during the collision of an atom and an atomic ion such that an electronically excited molecular ion is formed. 
	}
	\label{Fig2}
\end{figure}

Various strategies for the production of cold molecular ions have been devised. A selection of molecular-ion generation processes is shown in Fig. \ref{Fig2}. Some of these enable the direct generation of molecular ions in well-defined quantum states with high fidelity. Examples are photoassociation \cite{Jones2006,Zuber2022}, and threshold photoionisation, where the ionisation process can be tailored in order to populate a specific energy level in the ion such as the rovibrational ground state \cite{Tong2010, Zhang2023}. In general, the formation of translationally cold molecular ions is facilitated by starting with initially cold atoms, molecules or ions. However, molecular-ion production is typically combined with subsequent schemes for external cooling, internal cooling and state preparation. A well-established technique to remove external or internal excitation is sympathetic cooling (or buffer-gas cooling) where the molecular ion collides with cold particles and subsequently thermalises. These cold particles could be laser-cooled atoms or atomic ions \cite{Molhave2000, Rugango2015, Wan2015,Rellergert2013, Hansen2014}. The cold ions then localise in the trap to form ordered structures often termed "Coulomb crystals" \cite{Molhave2000}. Recently, forced evaporative cooling of a molecular anion with the help of photodetachment was also demonstrated \cite{Tauch2023}.  Typically, molecules are readily prepared in their vibrational ground state. In order to prepare the molecules also in well-defined rotational and even hyperfine states, powerful cooling and preparation techniques based on optical pumping have been implemented \cite{Staanum2010, Schneider2010, Lien2014, Stollenwerk2020, Bressel2012}.

\begin{figure}[t]
	\includegraphics[width=\columnwidth]{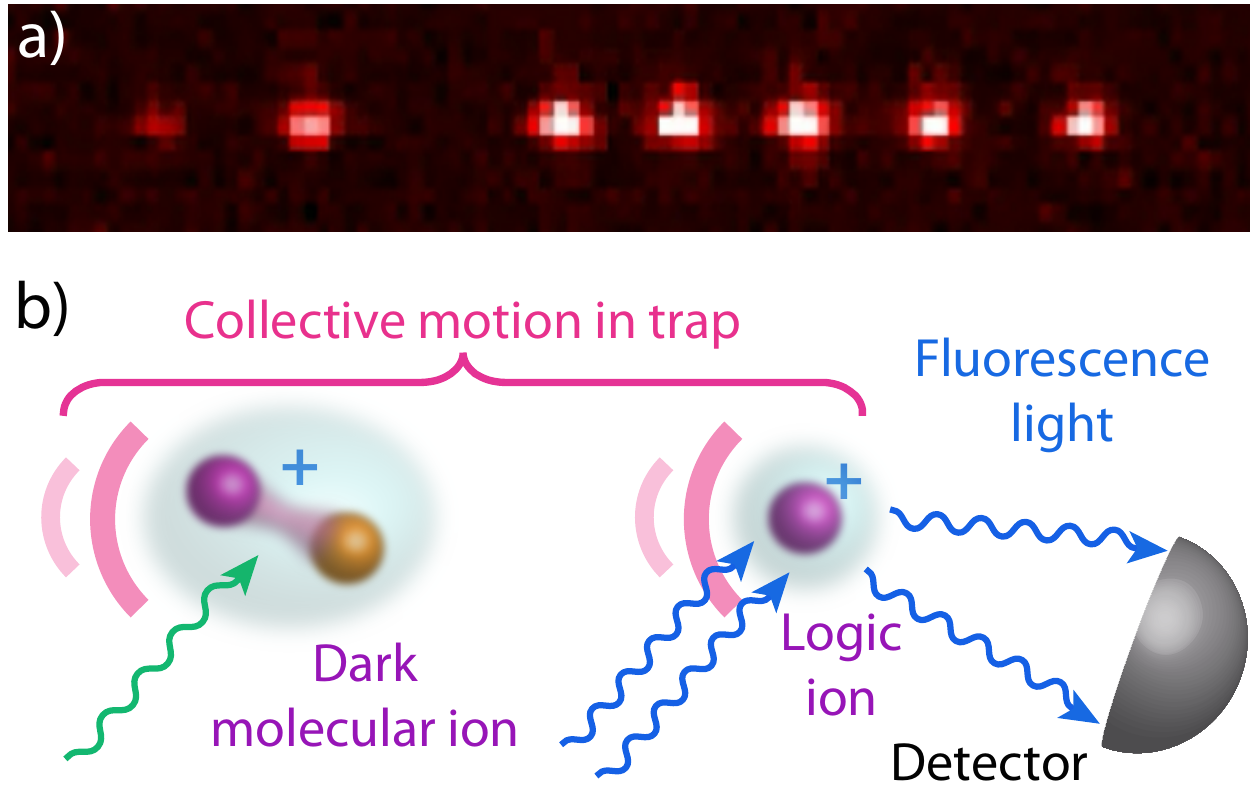}
	\caption{(a) Fluorescence image of an ion string in a linear Paul trap. The bright spots are atomic ions that scatter light of the detection laser. The dark, non-fluorescing ion at the third position from the left could, e.g., be a molecular ion. (b) Illustration of molecular-state detection using quantum logic. Information about the internal state of the molecular ion is mapped on the collective motion of both ions in the common trap. This mapping can be realised by, e.g., state-dependent optical forces or transitions (indicated by the green arrow). The collective motion is subsequently measured by optically probing the atomic logic ion and collecting its fluorescence (see blue arrows).
	}
	\label{Fig3}
\end{figure}

Sensitive methods have also been developed for the detection and analysis of trapped molecular ions. Laser-cooled Coulomb-crystallised atomic ions in the trap can readily be imaged and spatially resolved using their resonance fluorescence [see Fig. \ref{Fig3} (a)]. In these images sympathetically cooled molecular ions appear dark in the crystal.  The chemical identity of the ions is usually probed using mass spectrometry. In the context of ion-trap experiments, non-destructive techniques, e.g., resonant excitation of the ion motion \cite{Drewsen2004, Fan2021}, or destructive methods, e.g., time-of-flight mass spectrometry following the ejection of the ions from the trap \cite{Schowalter2012,Roesch2016, Schmid2017}, are frequently employed.

Spectroscopy of the trapped molecular ion and probing its internal quantum state is typically done using methods including laser-induced fluorescence, resonance-enhanced multi-photon dissociation and laser-induced reactions \cite{Willitsch2011, Sinhal2023,Calvin2023, Khanyile2015}. The drawback of these traditional techniques is their invasive and often destructive nature. Here, the internal quantum state is changed during excitation, and the molecule is often destroyed or chemically altered as a result of the detection. As a consequence, the sensitivity of these approaches is typically limited, in particular when working with small ion numbers.

As an alternative, quantum-logic spectroscopy has recently emerged as a potentially non-invasive and non-destructive technique to probe single trapped ions \cite{Schmidt2005, Sinhal2023}. Several variations of this approach have been developed for molecular species \cite{Wolf2016,Chou2017,Sinhal2020, Campbell2020}. All of these rely on the coupling of a molecular to an atomic ion in the trap [see Fig. \ref{Fig3} (b)] and the transfer of information about the state of the molecule to the atom via, e.g., state-dependent forces \cite{Wolf2016,Sinhal2020} or optical transitions \cite{Chou2017}. The information on the state of the molecule is read out afterwards from the atom using state-dependent resonance fluorescence. The advantage of this approach is its exquisite sensitivity
and that it enables probing in a quantum-non-demolition manner \cite{Wolf2016, Sinhal2020}. Furthermore,  quantum-logic spectroscopy can also be employed for the preparation of molecular ions in well-defined quantum states. Here, the preparation is based on the quantum mechanical projection process during the measurement \cite{Chou2017, Najafian2020}. On the down side, current implementations of quantum-logic spectroscopy are highly molecule specific and not suited for recording broad overview spectra.

\section{Ion trapping and micromotion}
\label{sec:MicoMotion}

\begin{figure}[t]
	\includegraphics[width=\columnwidth]{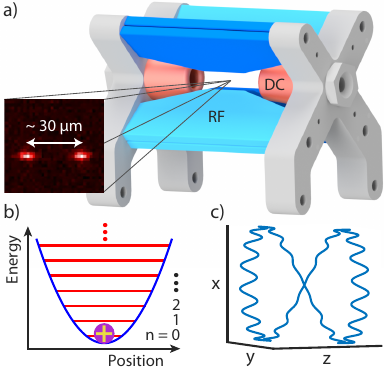}
	\caption{(a) Schematic of a typical linear Paul ion trap consisting of a combination of RF electrodes for radial and electrodes with DC voltages for axial confinement. The inset shows a fluorescence image of two atomic ions confined in the central region of the trap. (b) Quantised motional states of an ion in a harmonic trapping potential. The purple filled circle marked by the "+" indicates an ion in the motional ground state, $n = 0$. (c) Trajectory of a trapped ion characterised by its slow secular motion superimposed by fast micromotion.
	}
	\label{Fig4}
\end{figure}

For the experiments considered in this review, the molecular ions are typically trapped in a Paul trap which operates with radiofrequency (RF) electric fields \cite{Berkeland1998} [see Fig. \ref{Fig4} (a)]. Working with a Paul trap can be very convenient as it features tight confinement, so that a single ion can be located and moved around with nm precision. Furthermore, the trap typically exhibits a depth on the order of 1 eV, so that an ion generally remains captured even after strongly exothermic reactions. Such an exothermic reaction could be, e.g., the formation of a molecular ion out of an atomic ion and a neutral atom, or it could be the light-induced dissociation (i.e. photodissociation) of a molecular ion. With the help of laser and sympathetic cooling, atomic and molecular ions can be prepared in the quantum mechanical motional ground state of the trap potential [see Fig. \ref{Fig4} (b)]. This is often a prerequisite for the aforementioned quantum-logic schemes, and it also allows for high-precision spectroscopy, as will be discussed further below.

In addition to the quantised secular motion in the harmonic trap potential, the ion
also exhibits micromotion. Micromotion is coherently driven by the fast ($\sim$ MHz) oscillating electrical fields of the Paul trap, see Fig. \ref{Fig4} (c). While the effects of micromotion can often be neglected in pure ionic ensembles, it becomes very relevant in mixtures of ions and neutral particles. Here, micromotion energy can be converted into disordered thermal motion through collisions. This gives rise to interesting, counter-intuitive effects. For example, by immersing a trapped ion into an ultracold cloud of atoms, the ion is not simply sympathetically cooled to the temperature of the atoms as expected from simple thermodynamical considerations. By contrast, the ion can also be heated, even far above the temperature of the atom cloud. At that point, the ion's secular kinetic energy does not obey a Maxwell-Boltzmann distribution but exhibits a power-law high-energy tail \cite{DeVoe2009,Meir2018,Rouse2017}. The distribution quite strongly depends on the mass ratio of the ion and the neutral particle as well as the experimental boundary conditions \cite{Cetina2012,Rouse2019, Rouse2018}. The larger the mass ratio, the smaller the heating effect (see also discussion in the next section).

Finally, we mention that tight trapping affects the collision physics. For this, we first consider the collision of an atom with an atomic ion in free space. The two collision partners can only form a molecular bound state if there is a way to release the binding energy. This energy release could occur via the emission of a photon, but this process is typically slow. As a consequence, the formation of bound states by two-body collisions is often suppressed (see, however, the section about chemical reactions further below for exceptions). But, if the ion is confined in a trap, the trap can temporarily absorb momentum and energy of the two-particle system which facilitates the temporary formation of a molecule \cite{Pinkas2023, Hirzler2022b}. Evidence for this process was recently observed in experiments \cite{Pinkas2023}.

\section{Ultracold collisions and ion-neutral interactions}

A central topic for the field of molecular ions are atom-ion collisions. In such a collision, the molecular ion might be the collision partner or the product of a chemical reaction. The physics of ion-atom collisions  is often quite independent of the nature of the ion, particularly at long-range.

In general, collisions can be elastic, inelastic or reactive. Elastic collisions do not change the internal states of the colliding particles and hence can be used, e.g., for buffer-gas cooling of the ion in a cold cloud of atoms. Various types of inelastic collisions exist which manifest themselves in relaxation, excitation, and/or  spin-flips. Finally, reactions entail a change in the chemical nature of the colliding particles. 

For a simple description of the physical interaction, we now consider a collision between an ion and a neutral atom in the electronic ground state. At large distances, the corresponding interaction potential is attractive and scales as $ 1/r^4$ where $r$ is the distance between the two particles. This attraction originates from the electric field of the ion inducing an electrical dipole in the polarisable atom. The $1/r^4$ interaction is more long-range than the $1/r^6$ van-der-Waals interaction between neutral particles. As a consequence, scattering and reaction cross sections are typically several orders of magnitude larger for ions than for neutrals \cite{Idziaszek2009,Haerter2014,Tomza2019}. 

An important class of atom-ion collisions are Langevin-type processes. In a classical picture, the two colliding particles fall onto each other in a spiralling motion after they have passed the centrifugal energy barrier associated with the collision. At close distances, reactions and inelastic processes can then take place. The rate for classical Langevin collisions is independent of the collision energy or temperature \cite{Zhang2017}. At ultralow collision energies, however, the classical picture does not hold anymore. For instance, due to centrifugal barriers the particles can then only interact if their relative motion has vanishing quantised angular momentum, and all higher partial waves are 
"frozen out". This is called the $s-$wave regime. It allows for accurately controlling  collisions, e.g., via tunable Feshbach resonances. In recent years, such a control has revolutionised the field of ultracold neutral atoms and molecules \cite{Chin2010}. Therefore, it is desirable to reach the quantum regime, i.e., the regime of few collisional partial waves, also with ion-atom systems.

However, due to the long-range $1/r^4$ interaction between the ion and the neutral particle, the temperature needed to reach the $s$-wave regime is quite low. Depending on the mass ratio between ion and atom, typically required ion temperatures range between 100 nK (e.g. for Rb/Rb$^+$) and 100 $\mu$K (e.g. for Li/Yb$^+$) \cite{Cetina2012,Lous2022}. Reaching such temperatures in a Paul trap embedded in a neutral-atom environment can be challenging due to collisional heating effects as discussed in the previous section. Nonetheless, in a recent experiment in which a heavy trapped $^{171}$Yb$^+$ ion was immersed into a cold cloud of $^6$Li atoms the $s$-wave regime could be entered \cite{Feldker2020}. In another experiment in which a $^{138}$Ba$^+$ ion was immersed 
into a $^6$Li cloud cold enough temperatures were obtained so that Fesh\-bach resonances were observed \cite{Weckesser2021}.

Another possible approach to reach the $s-$wave regime is to use an optical dipole trap for trapping the ion, instead of a Paul trap, see e.g. \cite{Schmidt2020}. However, this method has the disadvantage that it leads to heating through spontaneous scattering of photons from the high-intensity laser fields. The high laser intensities are necessary because the optical dipole trap has to overcome the comparatively strong forces on the ion due to small uncompensated electrical stray fields which are typically on the order of a few mV/m.

Interestingly, it has been discovered that some scattering properties of the $s-$wave regime can even prevail at high temperatures when a large number of partial waves contribute to scattering \cite{Sikorsky2018,Cote2018,Katz2022b}. This effect occurs in resonant exchange processes such as spin exchange or charge transfer between the particles.
In general, scattering phases and cross sections of different partial waves are quite independent of each other. Thus, a mix of partial waves will generally average out the features of a single channel. In exchange processes, however, it is the difference of the scattering phases of two channels within a given partial wave which determines the cross section. These phase differences are fairly independent of the specific partial wave, resulting in similar exchange cross sections. Such a scenario has recently been observed for a spin-exchange process within a Rb/Sr$^+$ system \cite{Sikorsky2018} and for charge-exchange within a  Rb/Rb$^+$ system \cite{Katz2022b}.

Atom-ion interaction can be tuned, e.g., with the help of Rydberg states, as these are
highly electrically polarisable. For instance, it was observed that the collision cross section between an ion and a Rydberg atom can be orders of magnitude larger than the Langevin cross section between an ion and a ground state atom \cite{Ewald2019}. Furthermore, schemes have been proposed where coupling to Rydberg states or Rydberg dressing can be used to suppress atom-ion collisions at close range \cite{Wang2020, Secker2017}.

\section{Chemical reactions and inelastic processes}
\label{sec:Reactions}

\begin{figure}[t]
	\includegraphics[width=\columnwidth]{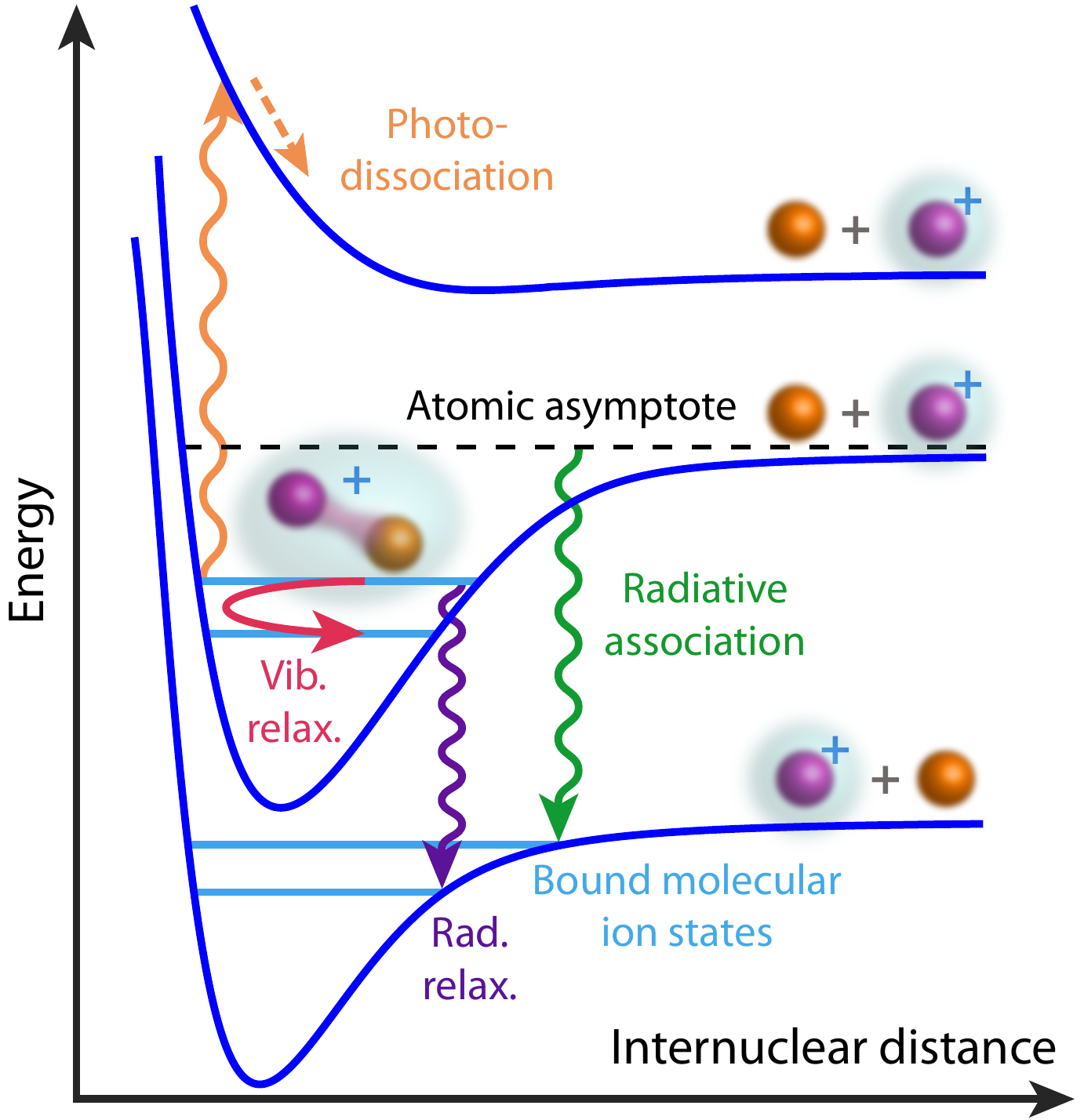}
	\caption{Scheme of molecular potential energy curves illustrating a selection of inelastic and reactive processes. Orange arrows: Photodissociation in which laser light couples a molecular bound state to a repulsive potential energy curve. Green arrow: Radiative association representing a molecular ion formation process. Purple arrow: Molecular radiative relaxation into another bound state of an energetically lower electronic potential. Red arrow: Vibrational relaxation which may be induced by a collision.
	}
	\label{Fig5}
\end{figure}

The study of chemical processes and reactions under precisely controlled experimental conditions has emerged as one of the key applications of cold molecular ions in traps.  
Among the first results from hybrid atom-ion trapping platforms was the demonstration of reactions that produce molecular ions from their atomic constituents. One such process is radiative association, where a molecular ion is formed by photon emission from the collision complex of the atom and atomic ion \cite{daSilva2015} (see Fig. \ref{Fig2} (b4) and Fig. \ref{Fig5}). This reaction was observed in early studies on ions interacting with alkali and alkaline-earth atoms \cite{Hall2011,Hall2013,Xing2022,Rellergert2011}. A second production process of molecular ions is three-body recombination [see Fig. \ref{Fig2} (b3)] where a dimer is formed in a collision of an atomic ion with two neutral atoms. In experiments with dense ultracold trapped atom clouds, this was found to be the dominant reaction mechanism \cite{Haerter2012, Kruekow2013a,  Kruekow2013b, Mohammadi2021}.
A third production process involves the reactive collision of a neutral molecule with an atomic ion [see Fig. \ref{Fig2} (b1)]. For example, in \cite{Hirzler2022a}, weakly-bound Li$_2$ Feshbach dimers collided with Yb$^+$ ions to produce LiYb$^+$ molecular ions. A fourth production process is photoionisation [see Fig. \ref{Fig2} (a1)] where, e.g., the dimer A$_2$ of an atom A is turned into an A$_2^+$ molecular ion after absorbing one or several photons with sufficient energy \cite{Haerter2013b, Jyothi2016, Hall2012}. Another process resulting in the generation of a molecular ion is photoassociation [see Fig. \ref{Fig2} (c)] \cite{Jones2006, Zrafi2020} which was recently observed \cite{Zuber2022,Zou2023}, creating a novel charged long-range Rydberg molecule \cite{Deiss2021,Duspayev2021}. Rydberg states also played an important role in yet another production process \cite{Zhelyazkova2020} where a cold collision between a neutral Rydberg atom and a neutral molecule led to the formation of a molecular ion. In somewhat related experiments, non-trapped molecular ions were generated within cold neutral beams using Penning ionisation [see Fig. \ref{Fig2} (a3)] \cite{Margulis2023, Gordon2018}.

\begin{figure}[t]
	\includegraphics[width=\columnwidth]{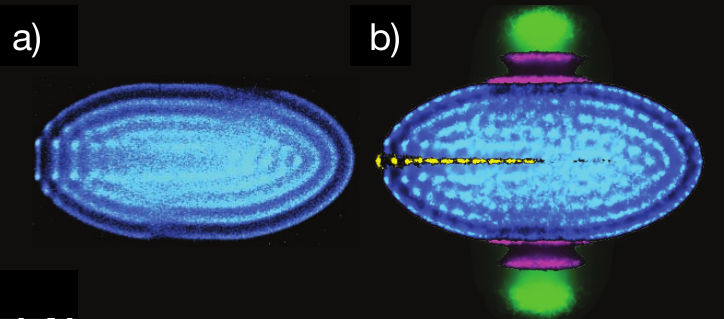}
	\caption{(a) Experimental and (b) simulated false-colour fluorescence images of Coulomb-crystallised laser-cooled Ca$^+$ ions recorded during the charge-transfer reaction of sympathetically cooled N$_2^+$ ions with cold Rb atoms. Sympathetically cooled ions appear dark in the experiment, but have been artificially made visible in the simulation. Colour code: yellow - N$_2^+$ reactant ions, green - Rb$^+$ product ions, magenta - CaH$^+$, CaOH$^+$ (products from side reactions). Adapted from Ref. \cite{Hall2012}. 
	}
	\label{Fig6}
\end{figure}

Once a molecular ion has been produced, it can undergo further reactive or inelastic processes. One example is the substitution of an atomic constituent of a molecular ion by a colliding atom \cite{Hall2011, Puri2017} [see Fig. \ref{Fig2} (b2)]. Another typical chemical reaction is charge exchange where an electron from a neutral collision partner is transferred to a molecular ion. As an example, the reaction N$_2^+$ + Rb $\rightarrow $ N$_2$ + Rb$^+$ \cite{Hall2012, Doerfler2019} was investigated within a cold atom-ion hybrid trap setup (see Fig. \ref{Fig6}). We note in passing that charge exchange processes between neutral atom - atomic ion systems have been commonly observed, see e.g. \cite{Grier2009, Schmid2010, Zipkes2010, Rellergert2011, Hall2011, Hall2013, Ravi2012, Mahdian2021}. In general, the charge exchange rates in a molecular collision depend on various parameters including the internal state of the particles. Studying these dependencies can give detailed insights in reaction pathways and provide control over the chemical reaction  \cite{Puri2019,Mills2019}. If the molecular ion is in an electronically excited state, a radiative decay towards an energetically lower state can occur. This could be radiative relaxation or even dissociation (see Fig. \ref{Fig5}) and can involve charge exchange between the atomic constituents of the molecule.

Furthermore, when optical light fields are present, photodissociation of the molecular ion can occur (see Fig. \ref{Fig5}). Light can drive a transition from a bound molecular state towards a repulsive potential curve, leading to a break up of the molecule with fragments of high kinetic energy \cite{Jyothi2016,Mohammadi2021, Hirzler2022a}.

Another interesting reactive process involving molecular ions is associative detachment, where a molecular anion A$^-$ and a neutral particle B combine to form a neutral molecule AB after emitting a loosely attached electron (see, e.g., \cite{Hassan2022} for results on the OH$^{-}-$Rb system). The emission of the electron is key for removing the binding energy, so that a stable bound state can be formed in a two-body mechanism with a potentially large cross section.

Besides chemical reactions, collisions between an energetically excited molecular ion and neutral particles can lead to relaxation processes where the molecular ion looses internal energy (see Fig. \ref{Fig5}). This has been observed in various setups for the vibrational and rotational degrees of freedom \cite{Rellergert2013, Hauser2015, Dieterle2020, Mohammadi2021} and used, e.g., for internal cooling \cite{Rellergert2013}. It can be interesting to study the relaxation paths within a molecule as they can potentially interfere and exhibit complex dynamics. Thus, in \cite{Mohammadi2021} the internal evolution of a highly excited  BaRb$^+$ molecule immersed into an ultracold cloud of Rb atoms was investigated.

While the first experiments on atom-ion platforms focused mainly on diatomic molecular ions, there has been recent work performed on polyatomic molecules \cite{Willitsch2008b, Voute2023, Puri2017}. Such experiments can offer valuable insights into complex reactive processes and reveal isomer- , conformer-, and stereospecific dependencies (see, e.g., \cite{Kilaj2018, Kilaj2021, Yang2021, Petralia2020}). For some of these studies, combinations of cold trapped ions with supersonic and effusive gas sources \cite{Willitsch2017} were used yielding collision energies well above 1K$\times k_\mathrm{B}$ and thus connecting to the astrophysical realm \cite{Krohn2022}.

\section{Precision measurements with molecular ions}

Their specific internal structure as well as their coupling to external environments can turn molecules into sensitive microscopic probes for fundamental physics. Molecules can be used, e.g., for testing the standard model or for searching for new particles and fields (see \cite{Safronova2018} for an overview that goes beyond the scope of the present work). High-precision spectroscopy is a powerful tool which gives access to the finest details of the molecular energy level structure in which the sought-after interactions might leave their traces. Trapped and cooled molecular ions are highly suitable for high-precision spectroscopy as observation times are potentially long and line broadenings (e.g. Doppler broadening) can be suppressed. This suppression is especially strong in the limit of tight confinement in which the Lamb-Dicke regime is reached \cite{Alighanbari2018}. 

A fundamental constant of interest is the proton-electron mass ratio. This ratio
has recently been measured via high-precision rovibrational spectroscopy of the hydrogen molecular ions  H$_2^+$, HD$^+$, D$_2^+$ \cite{Patra2020, Alighanbari2020, Kortunov2021}. These ions represent important benchmark systems as, due to their simplicity, \textit{ab initio} calculations of their level structure have reached an extraordinary level of accuracy. The proton-electron mass ratio was determined with a precision on the $10^{-11}$ level. 
	
Another fundamental parameter of interest is the electron's electric dipole moment (EDM). The magnitude of the electron EDM can constrain theories beyond the standard model which could explain aspects of dark matter and the matter-to-antimatter ratio in the universe \cite{Chupp2019}. So far, an EDM of the electron has not been observed. However, there is continuous progress in reducing upper-bound values. For such studies, one can exploit the large electric fields within polarised atoms or molecules as well as an enhancement of sensitivity due to the relativistic motion of a bound electron. A spin precession experiment using highly polar, heavy HfF$^+$ molecular ions \cite{Cairncross2017, Roussy2021} (see also \cite{Loh2013} for methods) has proven that an approach on the basis of trapped molecular ions can be competitive with other approaches using atomic or molecular beams with high particle flux. Limitations in statistical sensitivity due to the comparatively low ion number in the trap (up to a few hundred) are compensated by long interrogation times on the order of a second (see also \cite{Zhou2020}).
	
Very long interrogation times were also essential in a spectroscopic study on sympathetically-cooled N$_2^+$ ions, in which narrow electric-dipole-forbidden transitions were investigated for the first time in molecular ions \cite{Germann2014}. In general, such transitions play an important role for atomic and molecular clocks (see, e.g., \cite{Schiller2014}) and for the implementation of qubits \cite{Najafian2020}.

Another approach for high-resolution spectroscopy is based on  quantum-logic schemes \cite{Wolf2016, Campbell2020, Schwegler2023} as discussed earlier. Using such a scheme and employing a frequency comb, a spectral resolution of transition linewidths on the order of hertz and a precise determination of the EDM of a molecular ion was reported \cite{Chou2020, Collopy2023}.

\section{Perspectives}

The recent developments in the preparation and detection of atomic and molecular ions and the progress of experiments in hybrid atom-ion traps open up new opportunities for research. In the following, we briefly highlight a few examples.

\textit{Feshbach control:} After the first observation of atom-ion Feshbach resonances \cite{Weckesser2021}, the next important step could be to use them for a controlled 
association of molecular ions in a precisely-defined quantum state, e.g., by magnetic field ramping \cite{Chin2010}.  It is expected that this technique which proved extremely successful for neutral particles should also work for atom-ion systems.

\textit{State-resolved chemistry:} 
A key question associated with chemical reactions, molecular relaxations and
inelastic collisions is their dependence on the quantum states of the collision partners and the resulting state distribution of the products. For this purpose, one can extend quantum-logic spectroscopy techniques used for atomic ions (see e.g. \cite{Katz2022a}) for the use with molecular ions, as first results in Ref. \cite{Najafian2020} show. One could, for instance, measure the released reaction energy or measure particular state-dependent properties of the product molecule (e.g. its polarisability) and in this way identify the quantum state of the product molecule.

Alternatively, one could use resonance-enhanced multi-photon ionisation (REMPI) to state-selectively dissociate the molecular ion into two ionic fragments. This scheme could be analogous to a technique developed recently for neutral molecules \cite{Wolf2017,Liu2021}, potentially resolving even hyperfine levels.

\textit{Quantum information processing:} A potential appeal of using molecules  relies on their large number of internal degrees of freedom, which allows for implementation of diverse qubits and qudits \cite{Najafian2020}.
Quantum information processing with molecular ions is still in its infancy, but 
first steps have been taken. For example, as an important milestone, in \cite{Lin2020} entanglement between a molecular qubit and an atomic ion was demonstrated.

\textit{More complex molecular ions:}
In the future, one can adopt the large toolbox that has been developed and demonstrated with diatomic molecular ions to also investigate polyatomic species which possess, e.g., additional vibrational and rotational degrees of freedom \cite{Patterson2018, Schindler2019}. Here, quantum-logic spectroscopy might turn out to be useful as it can probe quasistatic properties of the molecule in a given quantum state, without the necessity of driving optical transitions. Simple polyatomic species are especially of interest for, e.g., the fields of astrophysics, atmospheric physics, and combustion. 

Beyond the examples discussed, many more routes for further research on molecular ions can be envisioned. It has been exciting to observe how the complementary approaches and platforms for studying ion-atom physics cross-fertilise and drive the development of the field. An increasing number of scientists are getting involved in this interdisciplinary field, and we are looking forward to a growing number of fascinating results.
\\

\noindent \textbf{Author contributions:} All authors contributed to the writing of the manuscript.
\\

\noindent \textbf{Competing interests:} The authors declare no competing interests.
\\

\noindent \textbf{Additional information:} Correspondence should be addressed to M.D. (markus.deiss@uni-ulm.de), S.W. (stefan.willitsch@unibas.ch), and J.H.D. (johannes.denschlag@uni-ulm.de).


%

\end{document}